\documentclass[10pt, conference, letterpaper]{IEEEtran}
\usepackage{epsfig}
\usepackage{amsmath}
\usepackage{amsthm}

\usepackage{subfigure}
\usepackage{graphicx}
\usepackage{cite}
\usepackage{amssymb}
\usepackage{fixmath}
\usepackage{caption}
\usepackage{algorithmic}
\usepackage{algorithm}
\usepackage[algo2e]{algorithm2e}
\usepackage[usenames]{color}
\usepackage{multirow}
\usepackage{flushend}
\usepackage{setspace}
\usepackage{soul}
\usepackage{pifont}

\usepackage{amssymb} 
\usepackage{tabularx}
\usepackage{makecell}
\usepackage{booktabs}

\usepackage[colorlinks=true, allcolors=black]{hyperref}

\usepackage{pifont}

\newcommand{\red}[1] {\textcolor[rgb]{0.00,0.00,0.00}{#1}}
\newcommand{\blue}[1] {\textcolor[rgb]{0.00,0.00, 0.00}{#1}}

\begin{document}

\IEEEoverridecommandlockouts

% \title{Janus: LLM-driven Unified Control and Intelligent Digital Twin for Dynamic Wireless Networks}

% \title{One Agent, Two Roles: Agentic LLM for Digital Twin Construction and Network orchestration}

\title{Janus: One Agent, Two Roles for Digital Twin Construction and Network Orchestration}

\author{
Peijun Hou, Yue Zhang, Yi Wei, Nan Cen\\
Department of Computer Science,
Saint Louis University, St. Louis, MO 63103\\
Email:{\{peijun.hou, yue.zhang.5, yi.wei.1, nan.cen\}@slu.edu
}
\vspace{-5mm}
}

\maketitle

\begin{abstract}
Next-generation wireless networks are increasingly large-scale, heterogeneous, and dynamic, posing significant challenges to network modeling and control. Digital twins offer a promising solution, but existing approaches typically rely on static models and decouple twin construction from network orchestration, limiting adaptability. Meanwhile, large language models (LLMs) provide strong reasoning capabilities but struggle with graph-structured data and are highly sensitive to prompt design, making inference unreliable.
In this paper, we propose \textbf{Janus}, an agentic LLM-based framework that unifies digital twin construction and network orchestration within a closed-loop system. 
At the core of Janus is a novel \emph{NetTwin agent}, which introduces a new \emph{one-agent, two-role} paradigm, enabling a single agent to simultaneously construct the digital twin from partial observations while performing task-driven network decision-making. To realize this, we design a modular architecture with three functional roles, including planner, modeler, and coder, along with a newly designed unified prompt module that further enhances graph reasoning robustness.
%Following a \emph{one-agent, two-role} paradigm, Janus enables a single agent to simultaneously construct and refine the digital twin from partial observations while performing task-driven network decision-making. To realize this, we design a modular architecture with three functional roles, including planner, modeler, and coder with a newly designed unified prompt module that further enhances graph reasoning robustness.
Extensive evaluations on wireless networks with up to 200 nodes and two representative tasks, including user association and SAGIN routing, show that Janus achieves over 90\% digital twin accuracy and near-optimal performance with gaps within 10.53\%, while maintaining moderate computational overhead. These results demonstrate the effectiveness, scalability, and generalizability of Janus, paving the way for AI-native autonomous wireless network orchestration.

\end{abstract}

\begin{keywords}
Wireless network, digital twin, agentic LLM, autonomous network orchestration
% Large language model, wireless network, digital twin
\end{keywords}

\IEEEpeerreviewmaketitle

% \end{itemize}}

%%%%%%%%%%%%%%%%%%%%%%%%%%%%%%%%%%%%%%%%%%%%%%%%%%%%%%%%%
%%%%%%%%%%%%%%%%%%% Introduction %%%%%%%%%%%%%%%%%%%%%%%%
%%%%%%%%%%%%%%%%%%%%%%%%%%%%%%%%%%%%%%%%%%%%%%%%%%%%%%%%%
\vspace{-2mm}
\section{Introduction}
\label{sec:intro}
\vspace{-2mm}

\noindent Next-generation wireless networks are becoming increasingly large-scale, heterogeneous, and dynamic, thus posing unprecedented challenges to network modeling and control. Driven by the rapid evolution from current 5G systems toward future 6G, wireless networks are expected to support ultra-high data rates, sub-millisecond latency, and ubiquitous connectivity across terrestrial, aerial, and space domains \cite{10855897}. These ambitious capabilities require the seamless integration of diverse communication technologies and ultra-dense network deployments, which significantly increase system complexity and operational overhead. As a result, traditional network control and optimization approaches struggle to effectively capture system dynamics and adapt to rapidly changing environments.

\begin{figure}[t]
\begin{center}
\includegraphics[width=0.3\textwidth]{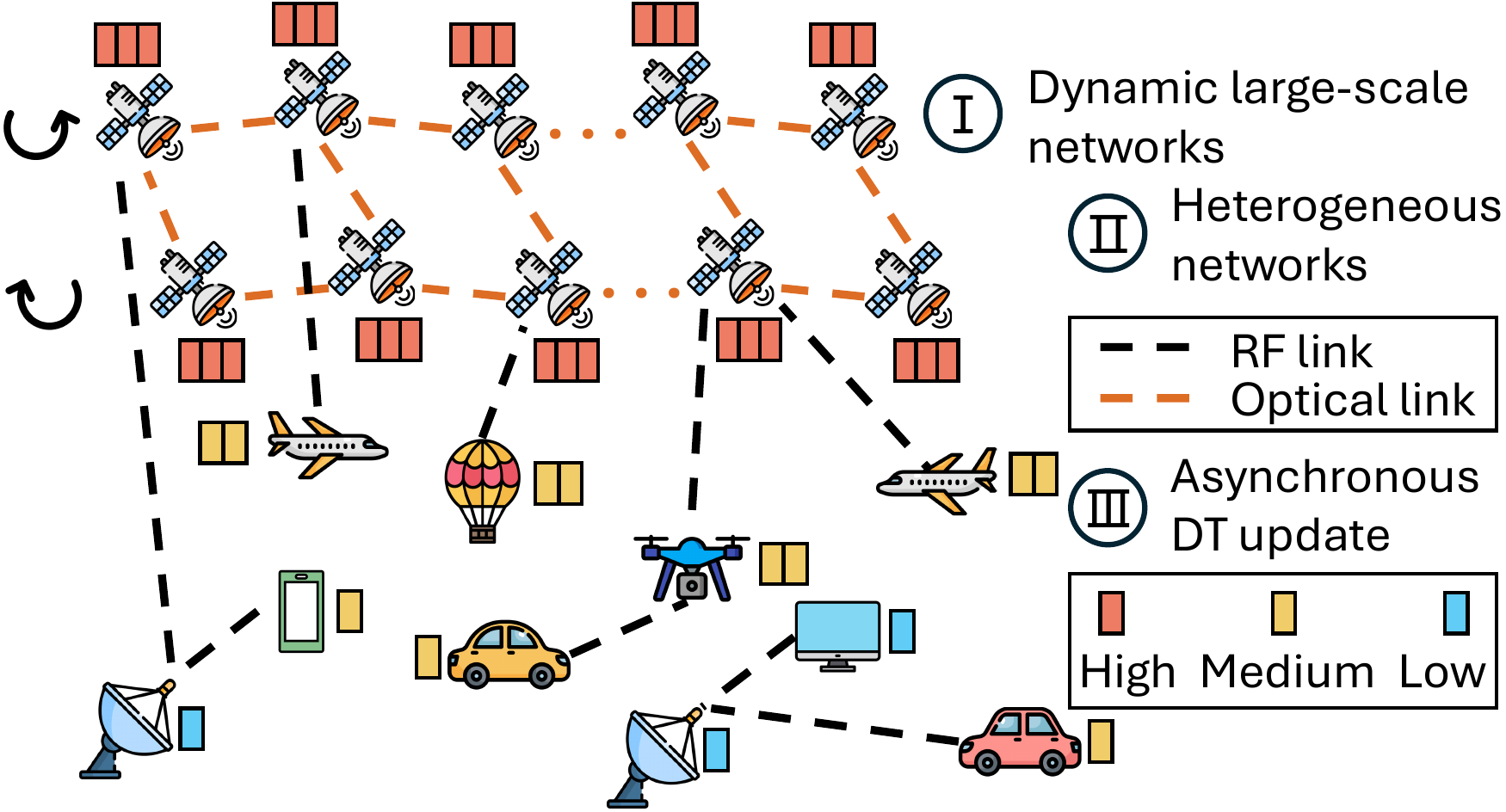}
\vspace{-3mm}
\caption{\small Dynamic large-scale wireless networks.} 
\vspace{-10mm}
\label{fig:scenario}
\end{center}
\end{figure}

To address these challenges, the concept of the \textbf{digital twin} (DT) has emerged as a promising paradigm for modeling and managing complex wireless systems. By maintaining a virtual representation of the physical network through continuous synchronization with real-time real-world networking data, digital twins enable system simulation, performance prediction, and resource optimization without disturbing live operations. In wireless networks, DTs provide a unified abstraction of multi-domain infrastructures, facilitate real-time monitoring, and support low-cost virtual simulation and experimentation. However, existing DT approaches face critical limitations. % \blue{as demonstrated in Fig.\ref{fig:scenario}}. 
In highly dynamic environments, \blue{as demonstrated in Fig. \ref{fig:scenario}}, heterogeneous communication latency often leads to incomplete or delayed data collection, making it difficult to maintain an accurate and up-to-date DT. Furthermore, learning-based DT models are typically trained under specific data distributions and may fail to generalize to unseen scenarios, limiting their robustness in real-world deployments.

Meanwhile, recent advances in large language models (LLMs) have demonstrated remarkable capabilities in reasoning, knowledge generalization, and handling incomplete information \cite{xu2024generate}. While initially designed for natural language processing, LLMs have evolved to effectively handle structured data, generate code, and enable complex decision-making across diverse domains \cite{11477808}. Emerging studies have also begun to explore their applications in wireless networking, such as resource allocation and routing \cite{zhou2024large, 11449190}. These developments suggest that LLMs can serve as a powerful foundation for next-generation AI-native networks, enabling intelligent network modeling, control, and decision-making.

However, existing efforts typically treat digital twin construction and network operation as two separate processes: the digital twin is built as a passive model \cite{hong2024llm}, while network control is performed by independent optimization or learning modules \cite{10592370}. This separation limits the ability of the system to adapt to dynamic environments and to fully exploit the synergy between modeling and decision-making.
%%%%%%%%%%%%%%%%%%%%%%%%%%%%%%%%%%%%%%%
%%%%%%%%%%%%%%%%%%%%%%%%%%%%%%%%%%%%%%%

%%%%%%%%%%%%%%%%%%%%%%%%%%
\begin{figure*}[t] 
\centering
\subfigcapskip=-3mm

\subfigure[Accuracy vs. number of nodes]{
\label{fig:pre_1}
\includegraphics[width=0.28\textwidth]{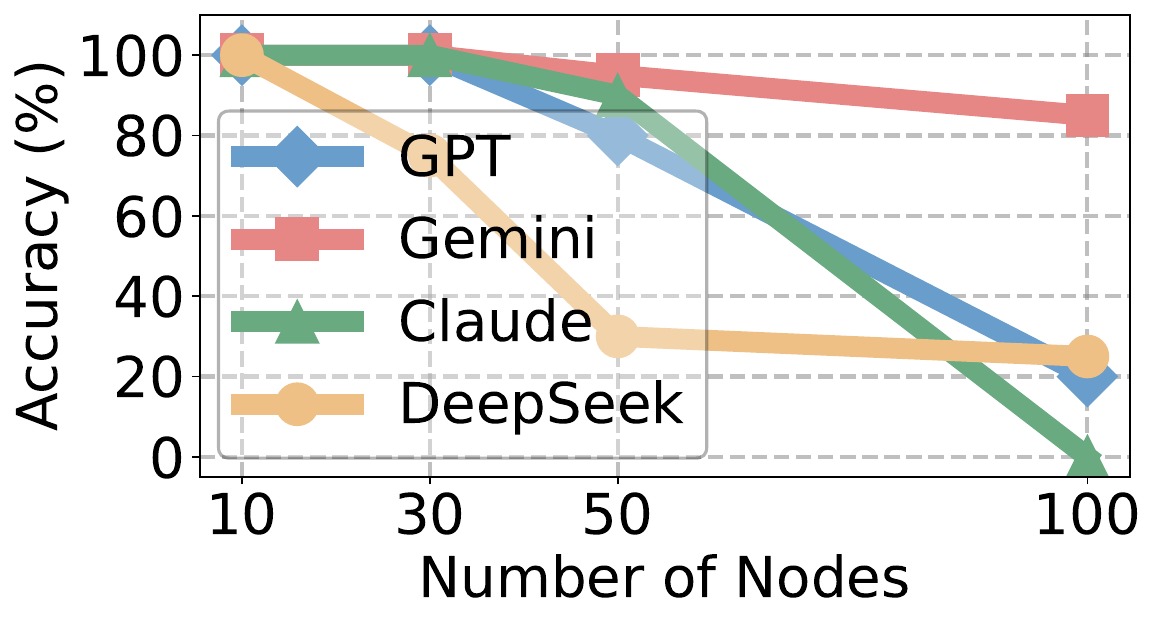}}
\hfill
\subfigure[Tokens vs. number of nodes]{
\label{fig:pre_2}
\includegraphics[width=0.27\textwidth]{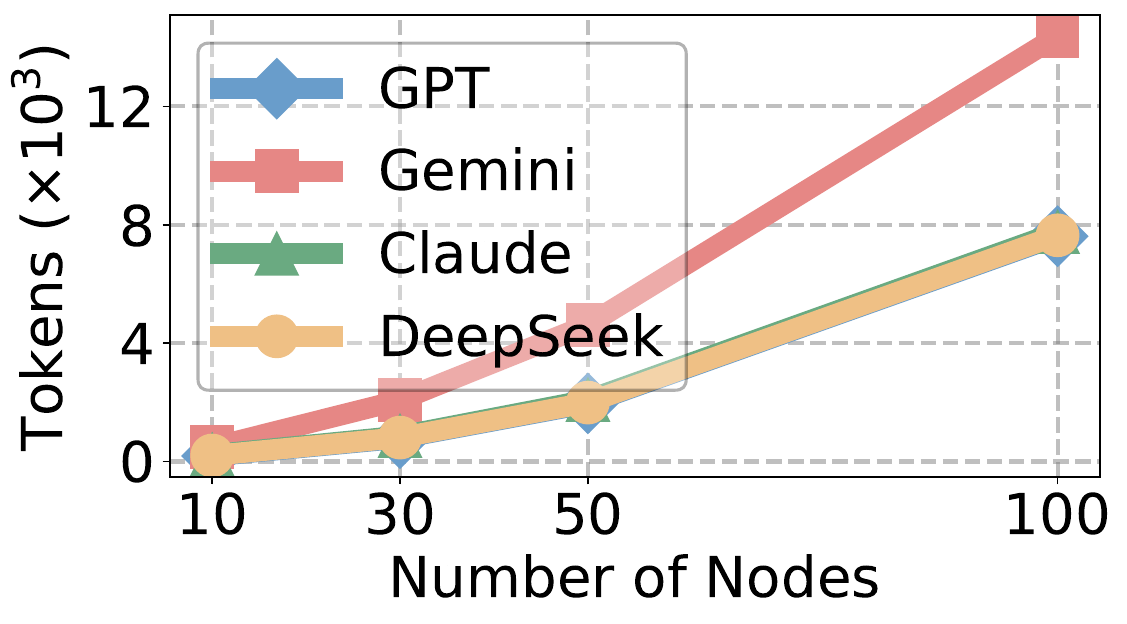}}
\hfill
\subfigure[Time vs. number of nodes]{
\label{fig:pre_3}
\includegraphics[width=0.269\textwidth]{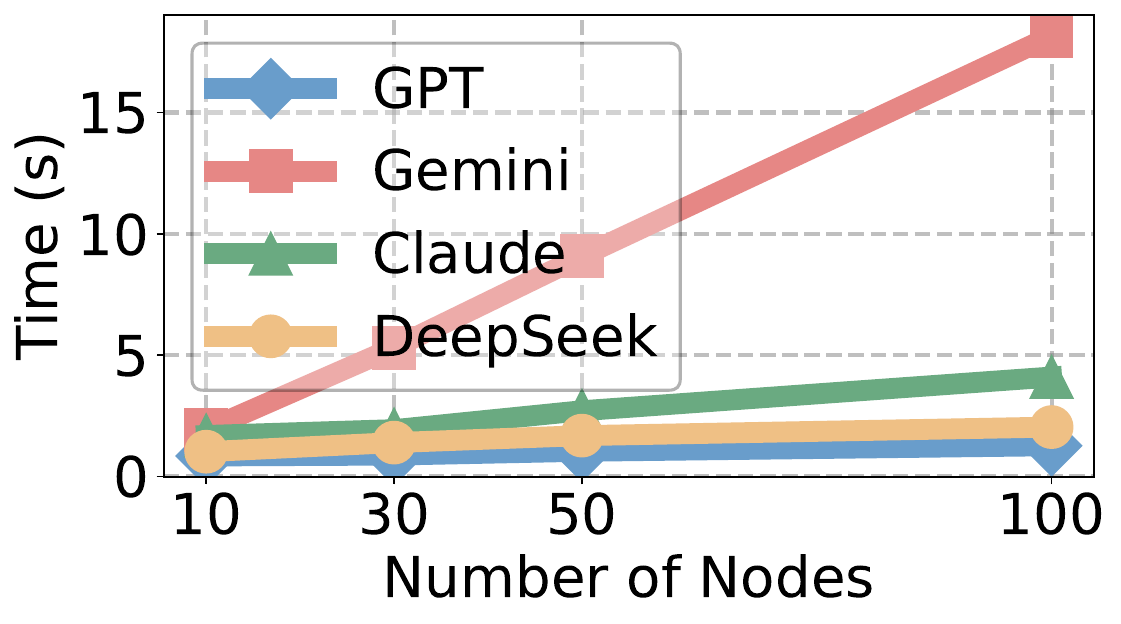}}
\vspace{-4mm}
\caption{\small LLMs for graph-structured data reasoning.}
\vspace{-8mm}
\label{fig:pre}
\end{figure*}
%%%%%%%%%%%%%%%%%%%%%%%%%%

In this paper, we propose \textbf{Janus}, a novel agentic LLM framework that unifies these two processes through a single agent. Inspired by the dual-faced nature of Janus, the proposed framework enables \emph{one agent to perform two complementary roles}. First, as a \emph{twin constructor}, the agent incrementally builds and continuously refines the network digital twin by integrating real-time networking data input and inferring missing information through its reasoning capability. Second, as a \emph{network orchestrator}, the same agent leverages the constructed twin to plan, evaluate, and execute network control and optimization in a closed-loop manner. By tightly integrating and coupling digital twin construction with decision-making, Janus establishes a self-evolving system that continuously improves both its internal model and operational performance.

The proposed Janus framework addresses key challenges in dynamic wireless networks. Its reasoning capability enables digital twin construction acceleration,
% robust operation under incomplete or delayed observations, 
while its strong generalization ability allows adaptation to unseen environments without fine-tuning. Moreover, by embedding digital twin construction and network control within a unified agentic work-loop, Janus enables continuous feedback and self-correction, leading to more efficient and reliable network management.

In summary, this paper makes the following contributions:

\noindent $\bullet$ We propose \textbf{Janus}, the first agentic LLM-based framework that unifies digital twin construction and network orchestration using a single autonomous agent.

\noindent $\bullet$ We introduce a ``one-agent, two-role'' paradigm and design a NetTwin agent to integrate DT construction with network control and optimization, enabling closed-loop interaction between network modeling and task-specific decision-making in dynamic and partially observable wireless environments. 
    %dual-role agent design that enables continuous interaction between modeling and decision-making.
    % \item  We develop a closed-loop architecture that integrates twin refinement and network control for dynamic wireless environments.
    
\noindent $\bullet$ We conduct comprehensive evaluations on representative network topologies and tasks, demonstrating that Janus achieves high accuracy, strong scalability, and near-optimal performance with moderate computational overhead. 
    %demonstrate the effectiveness of the proposed approach through comprehensive evaluation on representative network topologies and tasks.

\vspace{-2mm}
\section{Related Work and Preliminary Observations}
\label{sec:preliminary}

\noindent In this section, we first review related work on digital twins and LLM-based network intelligence. We then present key preliminary observations that motivate the design of Janus. Specifically, we focus on two critical aspects: (\textbf{i}) the limitations of frozen LLM in graph-structured data reasoning, and (\textbf{ii}) the impact of prompt design on LLM performance.
% \noindent In this section, we present key preliminary observations that motivate the design of Janus. Specifically, we focus on two critical aspects: (\textbf{i}) the limitations of frozen LLM in graph-structured data reasoning, and (\textbf{ii}) the impact of prompt design on LLM performance.

\vspace{-2mm}
\subsection{Related Work}
\noindent Recently, a growing body of work has explored the application of LLMs in wireless networks. In particular, several studies investigate the usage of LLMs in digital twin construction \cite{10829752, 10757470, duran2025gentwin}. For example, \cite{10829752} constructs a digital twin for optical networks to accurately capture physical dynamics and provide continuously updated network information to LLMs, thus enabling reliable decision-making. Similarly, \cite{10757470} proposes LINKs, an LLM-driven multi-agent framework that optimizes data retrieval and radio resource management for 6G digital twin networks. In addition, other works focus on LLM-based network intelligence. Some approaches leverage prompt engineering to enable in-context learning, guiding pretrained LLMs to solve specific wireless networking tasks, such as routing \cite{bao2025dynamic, 11449190}, resource allocation \cite{11358889}, and network intrusion detection \cite{10901312}. Other studies focus on fine-tuning LLMs to adapt them to domain-specific tasks, including channel-related tasks \cite{11071329} and network troubleshooting \cite{10.1145/3718958.3750505}. 

While these studies demonstrate the potential of LLMs and digital twins for wireless network intelligence, several fundamental challenges remain. In particular, effectively applying LLMs to wireless networks requires handling large-scale graph-structured data \cite{guo2023gpt4graph} and carefully designing prompts to ensure reliable reasoning \cite{sahoo2024systematic}. To better understand these challenges and their impact on system performance, we next present key preliminary observations that motivate Janus. % that motivate the design of Janus.

% \red{XXX write more XX}
\vspace{-2mm}
\subsection{Limitation of Frozen LLMs for Graph-Structured Data Reasoning}
\label{sec:sec:pre_graph}

\noindent Wireless networks are naturally represented as graph-structured data, where nodes represent network entities and edges capture their relationships. However, existing LLMs are primarily designed for sequential natural language processing, making them inherently inefficient for reasoning over graph-structured inputs \cite{guo2023gpt4graph}. A common approach is to encode graph structures into textual representations (e.g., edge lists), but this introduces significant scalability challenges.

To investigate this limitation, we conduct first-order neighbor identification experiments, which emulate essential networking tasks such as recovering routing tables. We evaluate four representative LLMs \blue{(GPT 5.4 mini, Gemini 2.5 flash, Claude Sonnet 4.6, and DeepSeek 3.2)} %\red{XXX including what XXX} 
on randomly generated graphs while varying the number of nodes. \blue{We evaluate each setting over 20 independent trials, and the LLM response is considered correct only if it exactly matches the ground truth.} Our results, \red{shown in Fig. \ref{fig:pre}}, reveal three critical trends.

% \red{XXX read the following three points and add corresponding figures for these two preliminariesXXX}

First, as illustrated in Fig. \ref{fig:pre_1}, \emph{accuracy degrades rapidly with increasing graph size}. While LLMs achieve high accuracy on small graphs (e.g., 10 or 30 nodes), their performance decreases significantly as the number of nodes increases, eventually \blue{approaching zero for larger graphs (e.g., 100 nodes) for GPT, Claude, and DeepSeek. Although Gemini shows a relatively smaller drop in accuracy, this comes at the cost of substantially higher resource consumption, requiring approximately twice the token usage and over 14$\times$ longer inference time compared to other models, as shown in Fig.~\ref{fig:pre_2} and Fig.~\ref{fig:pre_3}}.
%it requires approximately twice the token usage and over 14× longer inference time than other LLMs.}
% approaching zero for larger graphs (e.g., 100 nodes). 
These results indicate that LLM inference struggles to maintain reliable reasoning over large graph inputs.

Second, Fig. \ref{fig:pre_2} shows that \emph{token consumption grows substantially with graph size}. \blue{For example, as the graph size increases from 10 to 100, token consumption of Gemini increases substantially from 476 to 14,376.} %\red{XXX add some number here xxx}  
This rapid growth is due to the expansion of the textual representation of graph topology, which leads to significantly longer input sequences. Therefore, 
%As the number of nodes increases, the textual representation of graph topology expands rapidly, leading to longer input sequences. This 
this not only pushes the limits of LLM context windows but also exacerbates information dilution, making it difficult for the model to maintain and memorize critical structural relationships.

Third, \emph{wall-clock time increases %\sout{significantly} 
with input scale} as demonstrated in Fig. \ref{fig:pre_3}. In particular, for Gemini, the wall-clock time reaches as high as 18.15 seconds. This trend is as expected, as larger graph representations result in longer input sequences and increased computational complexity during inference, thereby leading to higher latency.
Such latency limits the applicability of LLM-based approaches in time-sensitive network control or orchestration scenarios.
%Such latency makes LLM-based approaches impractical for real-time network control or orchestration scenarios where timely decision-making is critical. %\red{XXX using digital twin will make it faster?XX }

These observations highlight a fundamental limitation: \emph {LLMs, particularly frozen LLMs, are not scalable, reliable, and efficient for graph-structured data reasoning}, especially in large and dynamic wireless networks.

%%%%%%%%%%%%%%%%%%%%%%%%%%
\begin{table}[t]
\footnotesize{
\centering
\vspace{+2mm}
% \begin{tabularx}{\textwidth}{X|X|X|X|X}
\renewcommand{\arraystretch}{1.1}
\begin{tabular}{c|c|c|c|c}
\hline
 % & \makecell{GPT} & \makecell{Gemini} & \makecell{Claude} & \makecell{DeepSeek} \\
 & GPT & Gemini & Claude & DeepSeek\\
\hline
Baseline & 43 &  87 & 67 & 27 \\
\hline
Query-first & 78 ($\uparrow$ 81\%) & 100 ($\uparrow$ 15\%) & 100 ($\uparrow$ 49\%) &  45 ($\uparrow$ 67\%) \\
\hline
\makecell{Role\\prompting} & 43 (-) & 97 ($\uparrow$ 11\%) & 79 ($\uparrow$ 18\%) & 61 ($\uparrow$ 126\%) \\
\hline
Few-shot & 44 ($\uparrow$ 2\%) & 63 ($\downarrow$ 28\%) & 76 ($\uparrow$ 13\%) & 49 ($\uparrow$ 81\%)\\
\hline
\makecell{CoT\\prompting} & 43 (-) & 91 ($\uparrow$ 5\%) & 100 ($\uparrow$ 49\%) & 58 ($\uparrow$ 115\%)\\
\hline
\makecell{Explicit\\instruction} & 57 ($\uparrow$ 33\%) & 94 ($\uparrow$ 8\%) & 89 ($\uparrow$ 33\%) & 60 ($\uparrow$ 122\%)\\
\hline
% \end{tabularx}
\end{tabular}}
\vspace{-2mm}
\caption{\small Reasoning accuracy vs. prompt design.}
%LLM accuracy on graph-structured data reasoning with different prompt design.}
\vspace{-8mm}
\label{tab:pre_table}
\end{table}
%%%%%%%%%%%%%%%%%%%%%%%%%%

\vspace{-2mm}
\subsection{Impact of Prompt Design on Graph Reasoning} \label{sec:sec:pre_prompt}

\noindent Beyond scalability limitations, we observe that LLM performance on graph-structured tasks is highly sensitive to prompt design. In wireless network digital twin construction, tasks such as missing information inference and topology reasoning require carefully structured interactions between the user and the LLM.
To systematically study this effect, we evaluate several prompt design strategies, including query ordering, role prompting, few-shot examples, chain-of-thought reasoning, and explicit instruction. 

\blue{We conduct the same first-order neighbor identification experiments with 30 nodes.} Our results \red{in Table \ref{tab:pre_table}} demonstrate that prompt design plays a crucial role in determining model performance. %\red{XXX same thing here, try to add number analysis XXX}
In particular, \emph{query-first and explicit instruction designs significantly improve accuracy across all evaluated LLMs}. \blue{Specifically, query-first achieves accuracy improvements of 81\%, 15\%, 49\%, and 67\% for GPT, Gemini, Claude, and DeepSeek, respectively. Similarly, explicit instruction leads to accuracy gains of 33\%, 8\%, 33\%, and 122\% for the same corresponding models.} These improvements highlight that placing the query before the graph description effectively guides model attention toward the task objective, while explicit instructions reduce ambiguity in task interpretation. In contrast, the effectiveness of role prompting and chain-of-thought reasoning varies across models, indicating their inconsistent and model-dependent effectiveness. Moreover, few-shot prompting may even degrade performance in certain cases, suggesting that additional examples do not necessarily translate into better reasoning for graph-structured tasks.
These findings reveal that LLM-based graph reasoning is not solely determined by model capability, but also heavily dependent on how the task is formulated. Consequently, \emph{prompt design becomes a critical factor for enabling reliable and efficient graph reasoning in wireless network applications}.

\textbf{In summary,} the above insights motivate the need for an agentic framework that goes beyond inference by incorporating structured reasoning, iterative refinement, and task-aware interaction, as detailed in the next section.

%%%%%%%%%%%%%%%%%%%%%%%%%%%%%%%%%%%%%%%
\begin{figure*}[t]
\begin{center}
\includegraphics[width=0.82\textwidth]{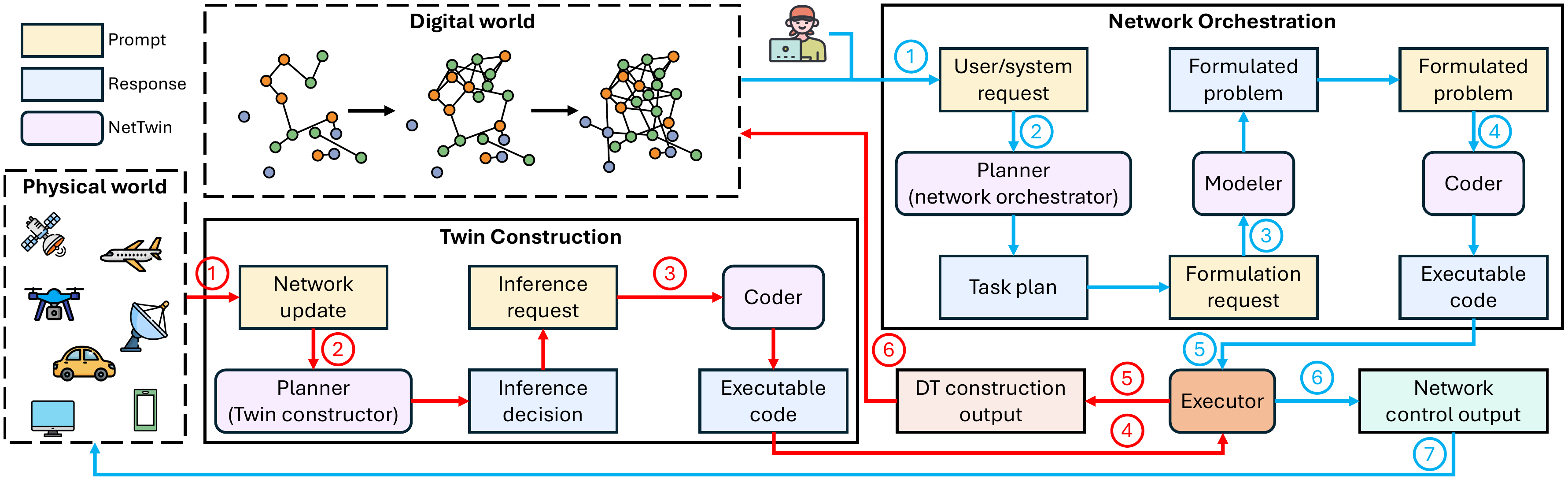}
\vspace{-2mm}
\caption{{\small The system overview and workflow of Janus.}} 
\vspace{-10mm}
\label{fig:diagram}
\end{center}
\end{figure*}

\vspace{-2mm}
\section{Janus Design}
\label{sec:design}
\vspace{-2mm}

\noindent Figure \ref{fig:diagram} illustrates the overall architecture and workflow of Janus. The goal of Janus is to provide a unified agentic framework that simultaneously enables (i) autonomous construction of wireless network digital twins based on incomplete and dynamic observations, and (ii) intelligent orchestration of diverse network tasks in a closed-loop manner. Specifically, Janus consists of two key modules: LLM-based \textbf{NetTwin Agent}, and \textbf{Unified Prompt Module}.  %\red{XXX please update if you have more XXX}

\vspace{-2mm}
\subsection{NetTwin Agent}
\label{sec:sec:agent}

\noindent Wireless networks are inherently graph-structured and highly dynamic, where nodes and edges convey complex spatial and functional relationships. However, as demonstrated in Sec.~\ref{sec:sec:pre_graph}, LLMs struggle to reason over graph-structured data when encoded into textual representations, where the performance of LLMs degrades rapidly as the network size increases. In addition, real-world wireless networks are only partially observable, with incomplete, delayed, and noisy observations. These challenges render static modeling and inference approaches insufficient for reliable network control and optimization. Moreover, effective network orchestration requires a closed-loop mechanism that continuously integrates new observations and execution feedback to adapt to dynamic environments. At the same time, such a system should support diverse and heterogeneous networking tasks, including resource allocation, routing, and fault diagnosis, within a unified framework. These requirements highlight the need for an agentic approach that tightly integrates dynamic network modeling with task-driven decision-making.

To address these challenges, we propose the \emph{NetTwin agent}, an agentic LLM-based framework that enables closed-loop digital twin construction and network orchestration without model fine-tuning. The NetTwin agent is composed of three core roles: \textbf{planner}, \textbf{modeler}, and \textbf{coder}, as shown in Fig.~\ref{fig:diagram}. Through coordinated interactions among these roles, the agent is capable of perceiving the environment, decomposing tasks, constructing and refining the digital twin, and executing network control decisions. Furthermore, the proposed design fundamentally transforms the system from a passive tool that merely responds to user prompts into an active agent capable of autonomously executing tasks and making decisions for wireless network digital twin construction and orchestration. In the following, we detail the core roles of the NetTwin agent.

$\bullet$ \textbf{Planner}.
The planner serves as the central controller of Janus, which is responsible for both environment perception and task decomposition. On the one hand, it analyzes the dynamic updates of the wireless network from the physical world and determines whether missing or unknown information can be inferred from the currently available observations. This capability enables the planner to guide digital twin construction and maintain a consistent and reliable network representation. 
On the other hand, the planner interprets user queries or system-level heterogeneous networking tasks and decomposes complex tasks into a sequence of structured sub-tasks. For example, in a dynamic space–air–ground integrated network (SAGIN), the planner determines whether routing tables require updates due to satellite mobility. 
It then decomposes the task into multiple steps, including problem formulation and executable code generation by the subsequent modeler and coder, respectively, thus ultimately producing updated routing decisions for the digital twin. Throughout this process, the planner selectively retrieves relevant information from the digital twin and aligns task objectives with the current network state. By integrating environment perception and task planning, the planner enables adaptive decision-making and continuous digital twin updates in dynamic and partially observable environments in a closed-loop lifecycle.

$\bullet$ \textbf{Modeler}.
To enable the LLM to gain a deeper understanding of the wireless network optimization problems, instead of solving them directly, we introduce a modeler to further analyze the problem by transforming the task into a formal optimization formulation, including sets, parameters, decision variables, the objective function, and constraints \cite{jiang2025llmopt}. 
For instance, a power control task can be formulated as an energy efficiency maximization problem subject to various budget constraints. This abstraction enhances interpretability and ensures that downstream reasoning is grounded in well-defined mathematical structures, thus producing more reliable solutions. Within the Janus framework, the modeler is primarily used to support network orchestration tasks.

% For example, it can formulate a power control task as an energy efficiency maximization problem with several budget constraints according to the user query, thereby providing a more reliable solution. In the Janus framework, the modeler is used for network orchestration only. 

$\bullet$ \textbf{Coder}. To address the inherent limitations of LLMs in mathematical computation, we introduce a coder that converts the reasoning outputs into executable code. Specifically, the coder translates the structured reasoning process into scripts (e.g., Python code), which are then executed by an external executor to obtain accurate results. This design bridges the gap between symbolic reasoning and numerical computation, significantly improving the reliability and correctness of the overall framework for both wireless network digital twin construction and network orchestration. 

% To address the inherent inaccuracy of LLMs in mathematical calculation, we introduce a coder that converts the reasoning process into the generation of executable code. This design ensures that the reasoning logic produced by the LLM is transformed into implementable scripts, such as Python code, which can then be executed by an external code executor to obtain precise numerical results. In this way, the coder enhances the reliability of the overall framework by ensuring accurate computation for wireless network digital twin tasks. 

%%%%%%%%%%%%%%%%%%%%%%%%%%%%%%%%%%%%%%%%%%%%
\begin{figure*}[t]
\centering
\subfigcapskip=-2mm

\subfigure[Physical world]{
\label{fig:topo_1}
\includegraphics[width=0.18\textwidth]{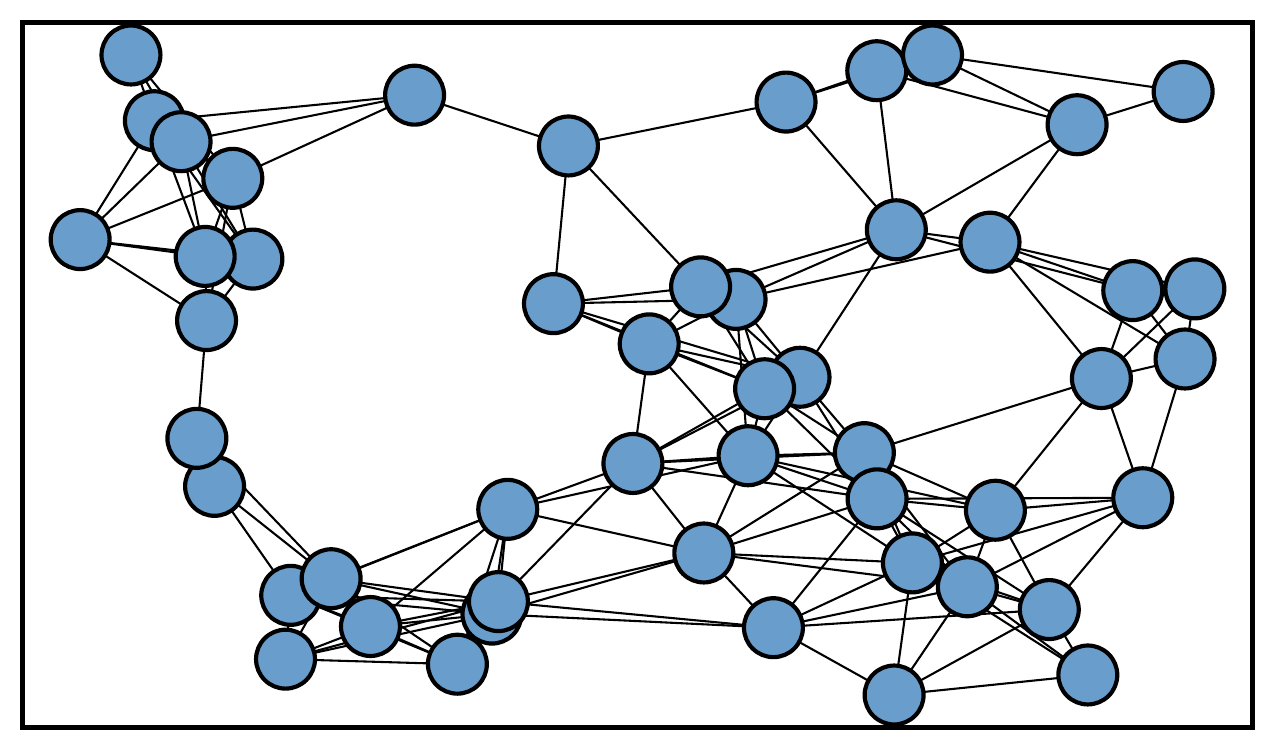}}
\hfill
\subfigure[DT topology at t = 0]{
\label{fig:topo_2}
\includegraphics[width=0.18\textwidth]{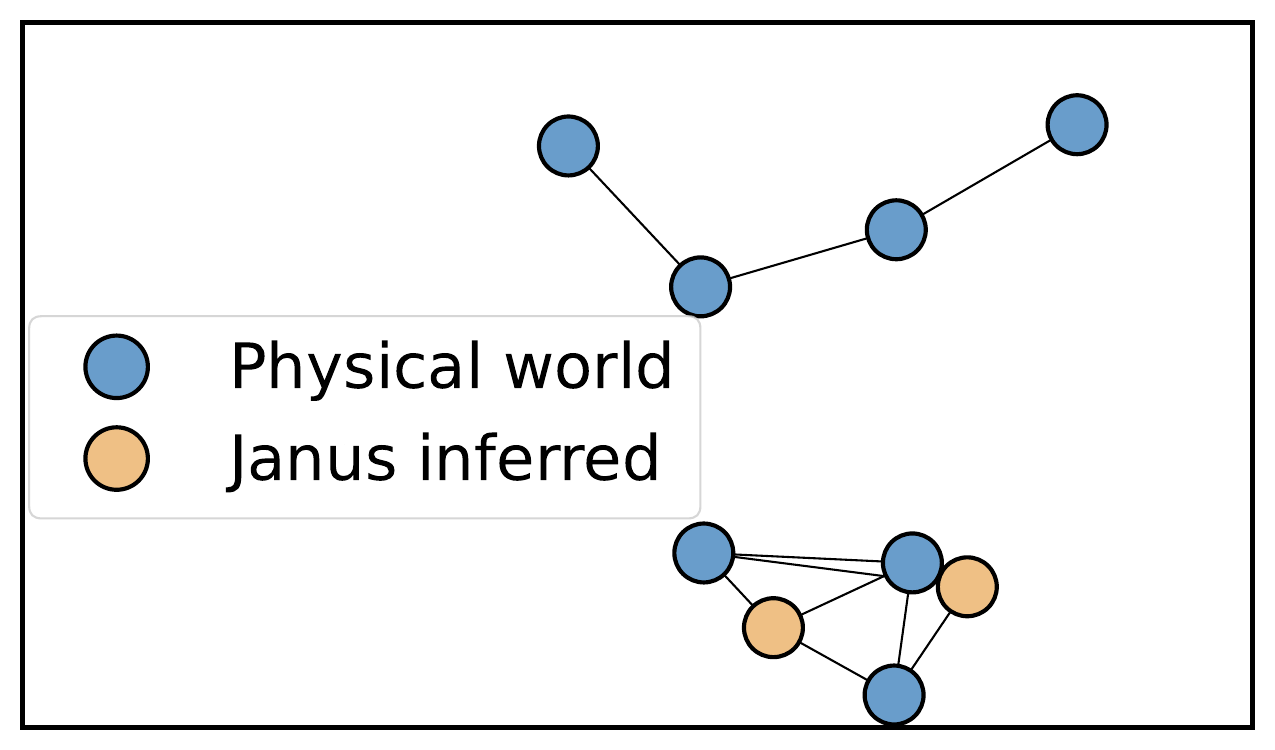}}
\hfill
\subfigure[DT topology at t = 4]{
\label{fig:topo_3}
\includegraphics[width=0.18\textwidth]{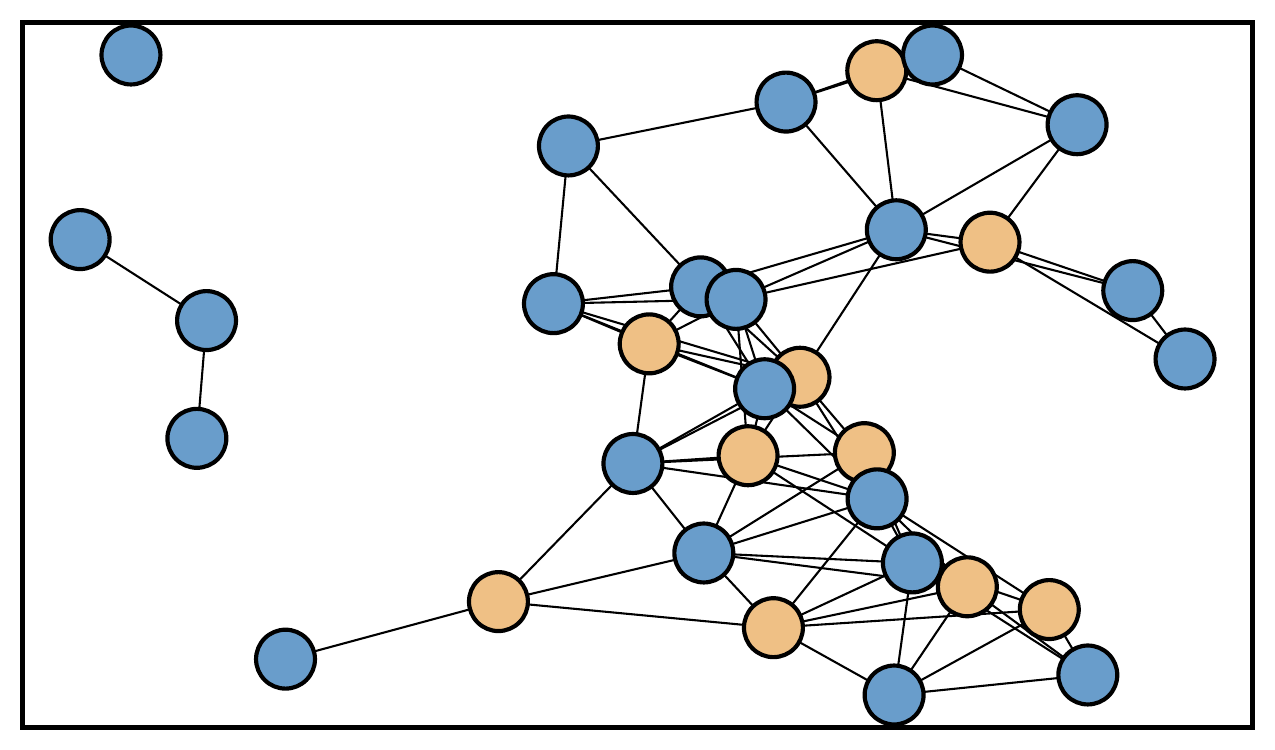}}
\hfill
\subfigure[DT topology at t = 8]{
\label{fig:topo_4}
\includegraphics[width=0.18\textwidth]{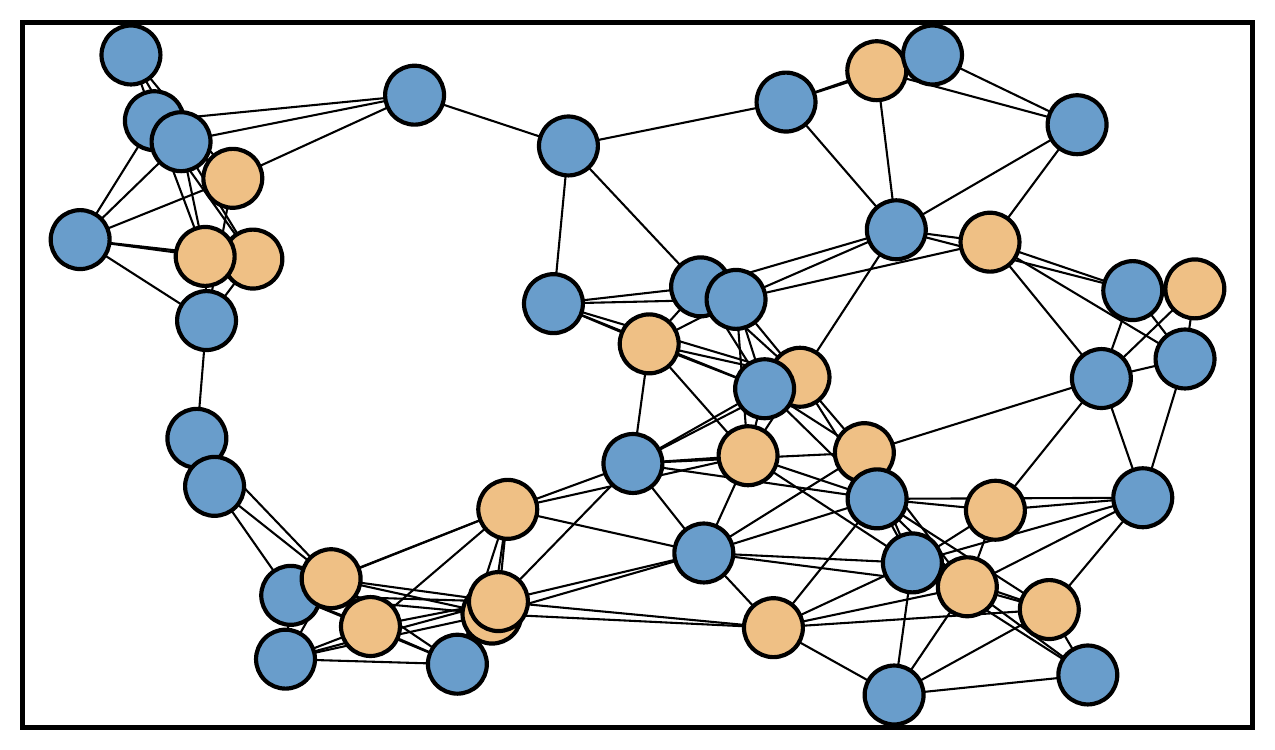}}
\vspace{-4mm}
\caption{\small Physical world topology and Digital Twin topology over time with Janus.}
\vspace{-8mm}
\label{fig:topo}
\end{figure*}
%%%%%%%%%%%%%%%%%%%%%%%%%%

\vspace{-2mm}
\subsection{Unified Prompt Module} \label{sec:sec:prompt}
\noindent As observed in Sec. \ref{sec:sec:pre_prompt}, LLM performance on graph-structured tasks is affected not only by %not solely determined by 
model capability, but also critically by %dependent on
how the task is formulated. Guided by these %empirical 
observations, % during our Janus framework development, 
we design a set of reusable unified prompt templates that systematically incorporate
key prompt engineering techniques to improve the performance of Janus. 

Specifically, our design integrates four widely used prompt engineering techniques \cite{sahoo2024systematic}. First, \emph{query-first} is used to place the core task query at the beginning of the prompt to guide the model’s attention toward the objective, such as placing instructions ``Your task is to solve the routing problem in a satellite-air-ground integrated network". Second, \emph{role prompting} is employed to assign a task-specific role to the LLM, e.g., ``You are a planning agent for a wireless network digital-twin system", thus aligning the model’s behavior with the desired functionality. Third, \emph{explicit instruction} is used to provide clear and detailed guidance on task objectives and constraints, such as ``Only use the provided digital twin state". Fourth, \emph{structured output} is used to enforce predefined output formats (e.g., JSON schemas) to ensure that model responses are well-structured and compatible with downstream modules.
% ensure the model responses are well-structured and compatible with downstream tasks by enforcing predefined JSON schemas. 
These unified prompt templates are tailored to different roles within Janus, including the planner (DT constructor and network orchestrator), the modeler, and the coder, as illustrated in Fig.~\ref{fig:diagram}. By standardizing interaction patterns across components, the proposed unified prompt module establishes a modular and extensible framework that significantly improves the robustness, interpretability, and generalization of the overall system.

% The unified prompt templates for the planner as a DT constructor, the planner as a network orchestrator, the modeler, and the coder are illustrated in Fig. \ref{}. These prompting techniques establish a modular framework that systematically improves the robustness, interpretability, and generalization of Janus. 

\vspace{-2mm}
\subsection{Agentic Workflow}

\noindent As shown in Fig. \ref{fig:diagram}, we next detail the workflow of two complementary roles of the proposed Janus framework, including (i) twin construction and (ii) network orchestration. 
%the agentic workflow of the proposed Janus framework, which includes two complementary roles: (i)twin constructor, which performs network state inference to construct and update the digital twin from incomplete and dynamic network environments, and (ii) network orchestrator, which plans, evaluates, and executes network management tasks in a closed-loop manner. Next, we introduce the agentic workflows in detail.

$\bullet$ \textbf{Twin construction}. 
The red path in Fig. \ref{fig:diagram} illustrates the digital twin construction workflow. The physical world periodically inputs partial and asynchronous updates, which are first organized into structured prompts via the unified prompt module (\ding{172}) and passed to the planner (\ding{173}). The planner determines whether the available information is sufficient to infer missing network states. If feasible, an inference request is sent to the coder, which generates executable code for state inference (\ding{174}). The code is then executed by an external executor (\ding{175}), and the inferred results are used to update the digital twin (\ding{176}-\ding{177}), enabling continuous refinement. 

% provides the agentic workflow of the digital twin construction in Janus, illustrating how the proposed Janus framework processes network updates. Specifically, the physical world periodically transmits partial and heterogeneous network updates. These updates are organized into structured prompts through the unified prompt module (\ding{172}) and then sent to the planner (\ding{173}). The planner then determines whether the newly obtained updates are sufficient to enable reasoning for inferring the missing or unknown network state. If the planner determines the inference is feasible, an inference request prompt is forwarded to the code to generate executable Python code for inferring the missing or unknown network state (\ding{174}). The generated code is then executed by an external executor (\ding{175}), producing the digital twin construction output (\ding{176}). Finally, the computed results are fed back into the digital twin, thereby continuously refining and improving the overall construction process (\ding{177}).

$\bullet$ \textbf{Network orchestration}. 
The blue path in Fig.~\ref{fig:diagram} shows the workflow for network orchestration. A task request, either from user queries or system-triggered events (e.g., routing updates due to satellite mobility), is first formatted into a structured prompt (\ding{172}). The prompt is passed to the planner, where it generates a task plan based on the current network state (\ding{173}), which is then passed to the modeler to formulate a formal optimization problem (\ding{174}). The coder converts this formulation into executable code (\ding{175}), which is executed by an external executor (\ding{176}) to obtain the solution (\ding{177}). The resulting decisions are lastly fed back to the physical network (\ding{178}), enabling effective network control and analysis.

% The blue line in Fig. \ref{fig:diagram} illustrates the agentic workflow for network orchestration, showing how Janus processes user queries or system-level heterogeneous networking tasks. In this workflow, a task request is first organized into a structured prompt (\ding{172}), either from a user query or triggered by system-level events (e.g., routing updates due to satellite mobility in SAGIN). The constructed prompt is then sent to the planner, which generates a task plan in response to user requests or when detected network changes require control actions (\ding{173}). The task plan is then forwarded to the modeler (\ding{174}), which formulates the task as a formal optimization problem and passes it to the coder (\ding{175}). The coder then translates the problem into executable code (\ding{176}), which is executed by an external executor to compute the network control solution (\ding{177}). Finally, the generated solution output is returned to the physical world (\ding{178}), enabling effective network control and analysis.

%%%%%%%%%%%%%%%%%%%%%%%%%%%%%%%%%%%%%%%%%%%%
\begin{figure*}[t]
\centering
\subfigcapskip=-2mm

\subfigure[Node completeness]{
\label{fig:dt_1}
\includegraphics[width=0.18\textwidth]{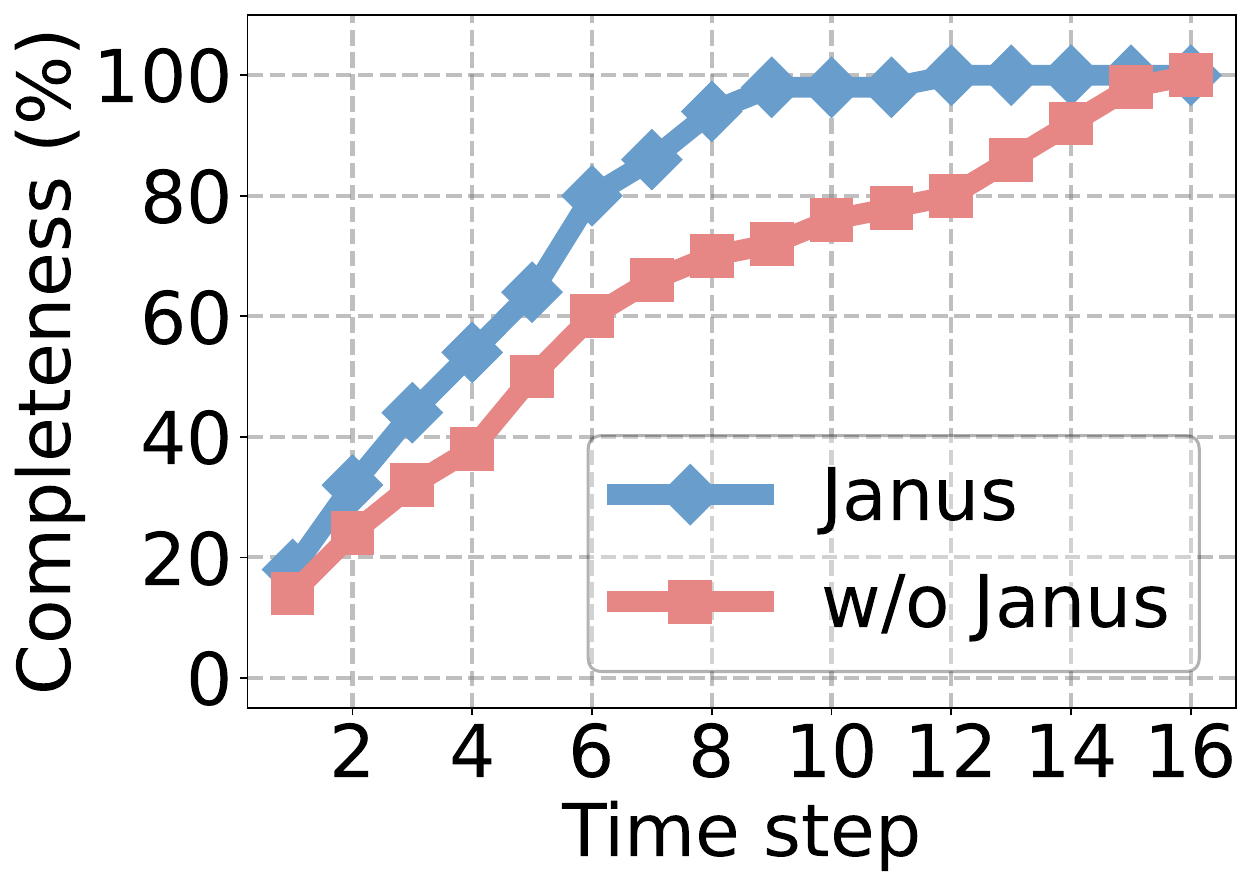}}
\hfill
\subfigure[Accuracy vs. number of nodes]{
\label{fig:dt_2}
\includegraphics[width=0.214\textwidth]{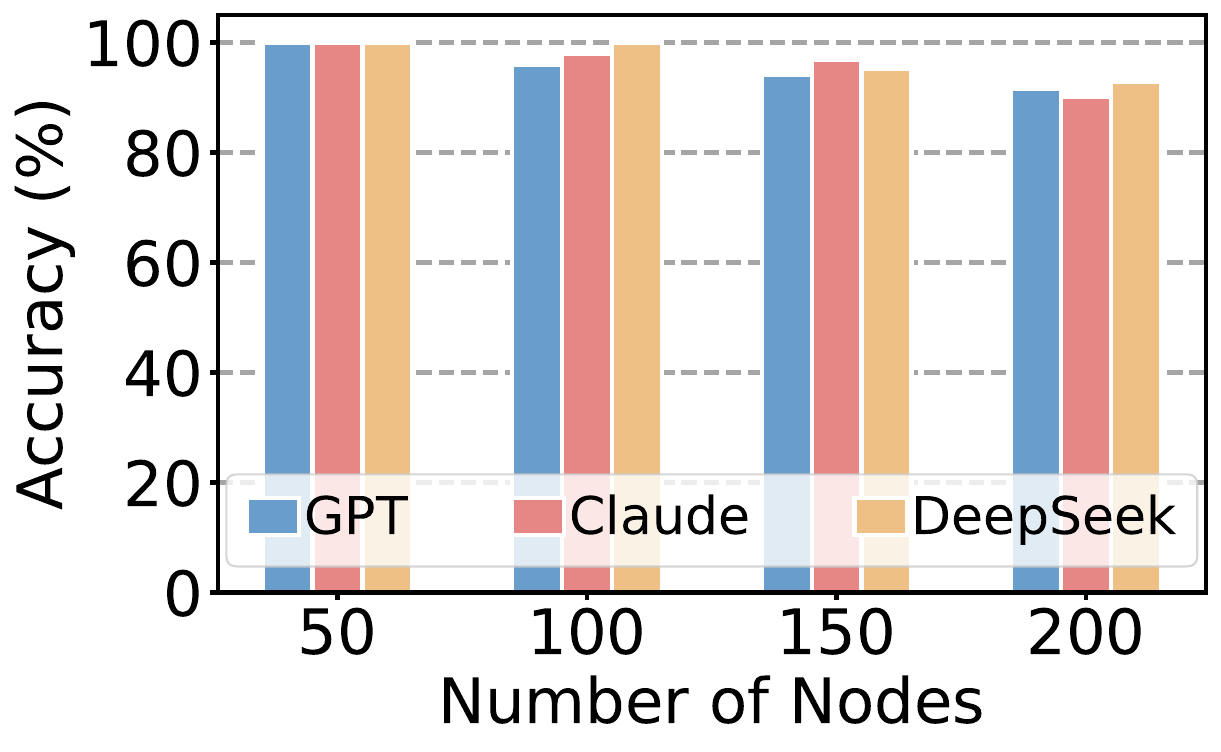}}
\hfill
\subfigure[Tokens vs. number of nodes]{
\label{fig:dt_3}
\includegraphics[width=0.2\textwidth]{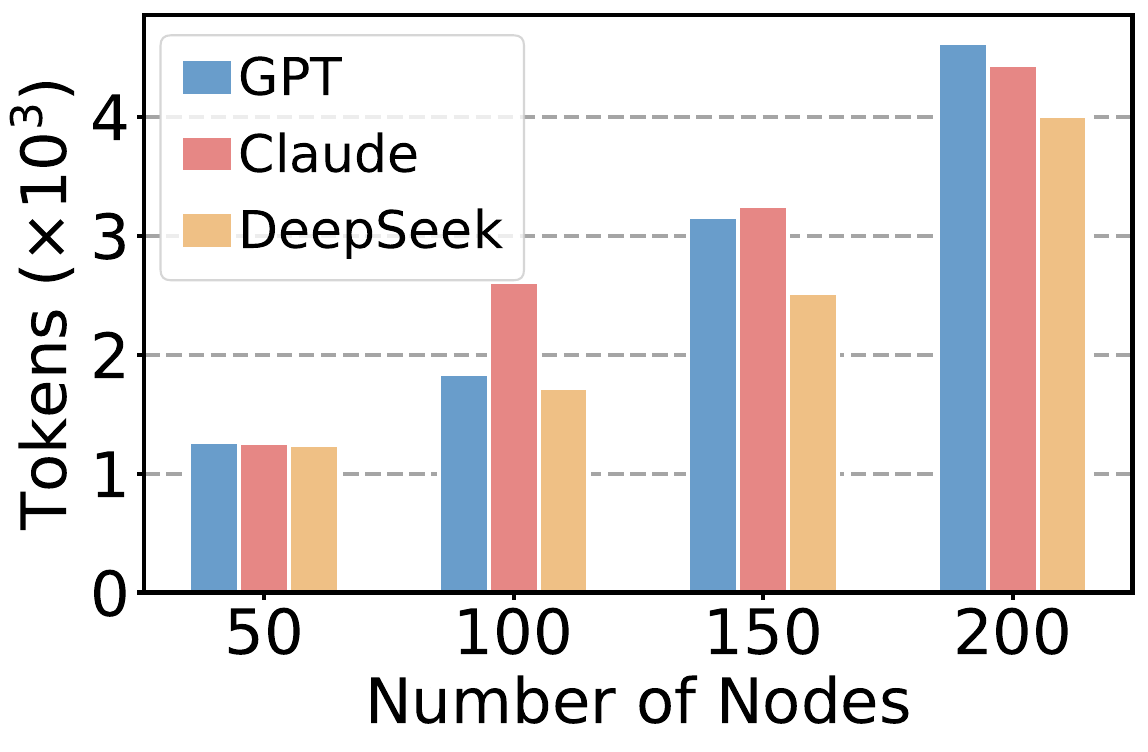}}
\hfill
\subfigure[Time vs. number of nodes]{
\label{fig:dt_4}
\includegraphics[width=0.206\textwidth]{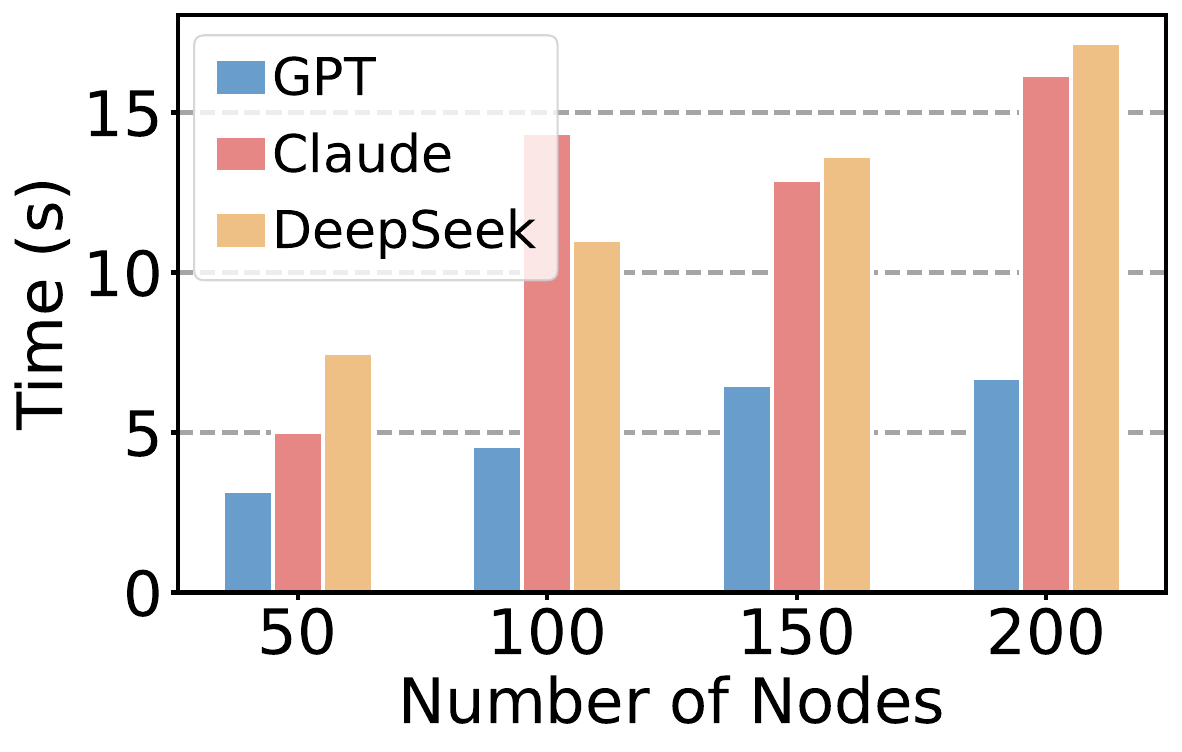}}
\vspace{-4mm}
\caption{\small The overall performance of Janus in digital twin construction.}
\vspace{-9mm}
\label{fig:dt}
\end{figure*}
%%%%%%%%%%%%%%%%%%%%%%%%%%

\vspace{-2mm}
\section{Performance Evaluation}
\label{sec:evaluation}
% \red{1-1.5 page}

\vspace{-1mm}
\subsection{Simulation Setup} \label{sec:sec:setup}

\noindent To evaluate the effectiveness of the proposed Janus framework, we consider an ad-hoc wireless network with various numbers of nodes. Specifically, for digital twin construction, we randomize the number of nodes from 50 to 200. At each time step, the physical network randomly provides partial node and edge information to the planner. Based on the available observations, Janus perceives the current state of the digital twin and infers missing but feasible information, such as node locations, thereby progressively refining the network twin representation.  Meanwhile, the planner simultaneously processes user queries related to network orchestration, enabling closed-loop interaction between the digital twin construction and task-specific decision-making. In addition, we consider two representative wireless networking tasks to evaluate Janus’s capability in handling dynamic environments and supporting network orchestration: 
%To evaluate the digital twin construction performance of the proposed Janus framework, we consider an ad-hoc wireless network with the number of nodes ranging from 50 to 200. At each time step, the physical world of the wireless network randomly provides node and corresponding edge information to the digital twin. Janus then perceives the current state of the digital twin and infers missing node information, such as node locations, based on the available network context. Furthermore, we consider two representative wireless networking tasks to evaluate Janus’s ability to perceive environmental dynamics and support network orchestration.
(i) \textbf{user association} in a 50-node indoor hybrid radio frequency (RF)/visible light communication (VLC) network, where the objective is %which aims 
to determine the optimal assignment between users and access points (APs) to maximize the network sum throughput subject to AP capacity and quality-of-service constraints. As users move within the indoor environment, Janus perceives the changes and dynamically updates the user-AP associations.
(ii) \textbf{routing} in a satellite-air-ground integrated network (SAGIN) based on the real-world 651-satellite OneWeb constellation \cite{eutelsat}, where the objective is to 
minimize end-to-end delay under link capacity and connectivity constraints. As satellite movement leads to intermittent link availability, Janus detects these changes and recomputes the routing path accordingly. 
%select paths between source and destination nodes to minimize the total end-to-end delay under link capacity and connectivity constraints. Space–ground links may become unavailable as satellites move. Janus identifies these changes and recomputes routing paths for the affected ground nodes accordingly.
Moreover, we implement Janus using three mainstream LLMs, including GPT-5.4 mini, Claude Sonnet 4.6, and DeepSeek 3.2.

To the best of our knowledge, there is no prior work that jointly addresses large-scale wireless network digital twin construction and network orchestration. Therefore, we implement a mathematical optimization-based benchmark, denoted as \emph{Optimal}. Specifically, each task is formulated as an optimization problem and solved using the Gurobi solver to obtain globally optimal solutions. While such approaches achieve optimality, they require task-specific formulations and lack generality. In contrast, Janus follows a unified ``one model for all'' paradigm, enabling a single agentic framework to handle diverse networking tasks. 

% targets the joint wireless network digital twin construction and network orchestration in large-scale wireless networks. Therefore, we implement a mathematical optimization-based benchmark, denoted as \emph{Optimal}. Specifically, the task is first formulated as a mathematical optimization problem and then solved using the Gurobi solver to obtain the globally optimal solution. While the mathematical optimization methods are capable of providing globally optimal solutions, they require different mathematical formulations tailored to each specific user task. In contrast, our proposed Janus framework provides a “one model for all” paradigm, enabling a single agentic workflow to address a wide range of wireless network tasks.

%%%%%%%%%%%%%%%%%%%%%%%%%%%%%%%%%%%%%%%
\begin{figure}[t]
\centering
\subfigcapskip=-2mm

\subfigure[User association in hybrid RF/VLC networks]{
\label{fig:use_case_1}
\includegraphics[width=0.45\columnwidth]{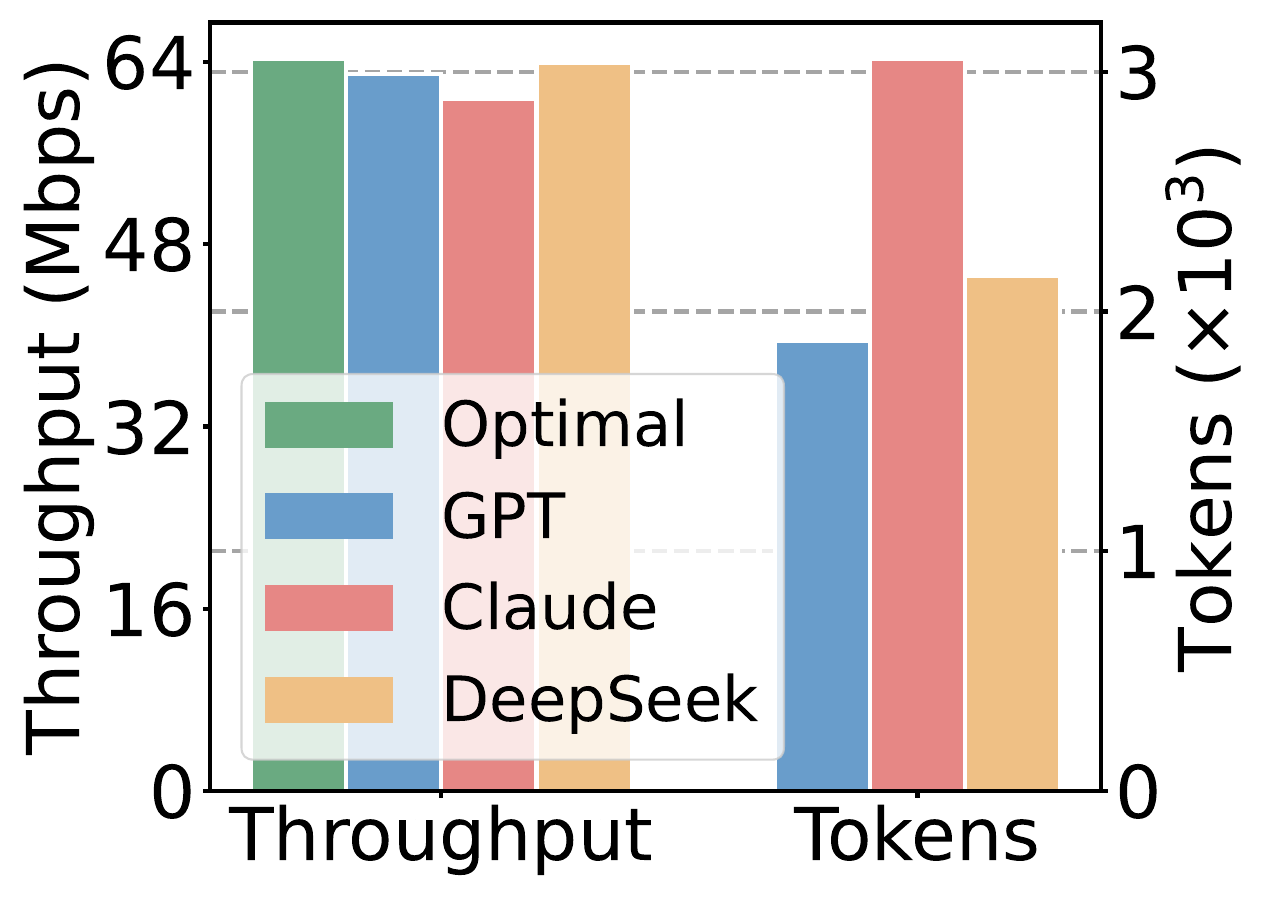}}
\hfill
\subfigure[Routing in OneWeb-constellation-based SAGIN]{
\label{fig:use_case_2}
\includegraphics[width=0.45\columnwidth]{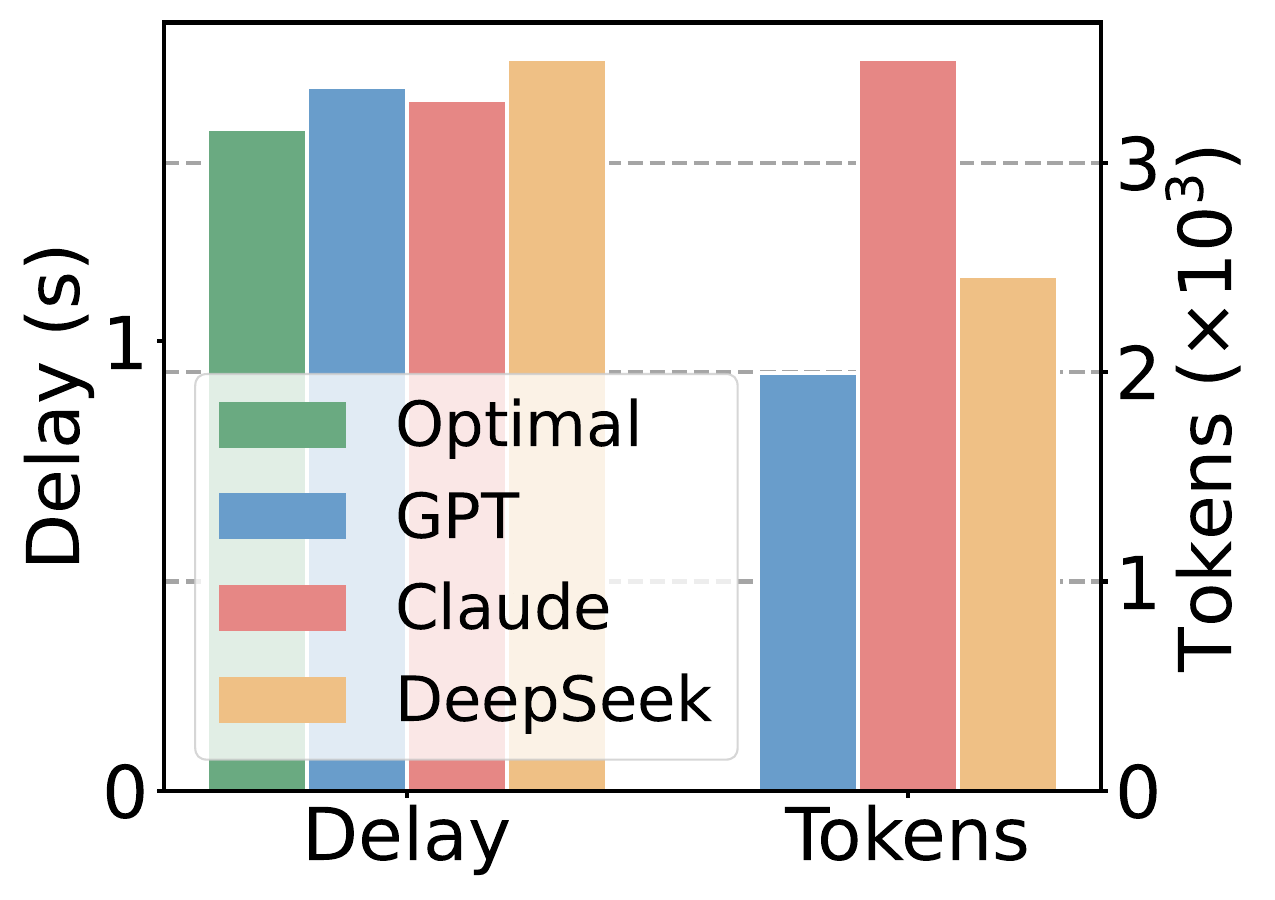}}
\vspace{-4mm}
\caption{\small Performance and token consumption for user association and routing tasks.}
\vspace{-8mm}
\label{fig:use_case}
\end{figure}
%%%%%%%%%%%%%%%%%%%%%%%%%%%%%%%%%%%%%%%

\vspace{-4mm}
\subsection{Evaluation Results}
$\bullet$ \textbf{Digital twin construction}. First, we evaluate the effectiveness of Janus as a digital twin constructor. 
Specifically, we begin with a 50-node network and compare the construction speed with and without Janus. Figure \ref{fig:dt_1} shows the results in terms of node completeness, which is defined as the ratio of nodes with complete information in the digital twin to the total number of nodes in the physical world. We can see that Janus achieves 100\% completeness within only 12 iterations, while without Janus, 16 iterations are used to fully reflect the physical world. 
%Without Janus, it takes 16 iterations for the digital twin to fully reflect the physical world, whereas Janus achieves 100\% completeness in only 12 iterations. 
To better understand the role of Janus in digital twin construction, Fig. \ref{fig:topo} compares the digital twin topologies at $t = 0, 4, 8$ with the ground-truth physical world topology. We observe that, as the construction process proceeds, an increasing proportion of node information in the digital twin is inferred by Janus. At $t = 8$, up to 19 out of 49 nodes are constructed through Janus, showing the effectiveness of Janus regarding the digital twin construction.

We further evaluate the scalability of Janus by increasing the network size up to 200 nodes. As shown in Fig. \ref{fig:dt_2}, Janus consistently achieves over 90\% accuracy across all evaluated LLMs, with only a slight degradation as the network size increases to 200 nodes. 
%the performance of Janus slightly degrades as the network size increases, as expected. Nevertheless, Janus consistently achieves over 90\% accuracy across all three LLMs, even when the network size reaches 200 nodes. 
In addition, as shown in Figs. \ref{fig:dt_3} and \ref{fig:dt_4}, Janus maintains reasonable computation overhead. For a network with 200 nodes, the wall-clock execution time is up to 6.72 s, 16.18 s, and 17.19 s for GPT, Claude, and DeepSeek, respectively, with corresponding token consumption of 
4626.65, 4439.20, and 4012.88. These results indicate that Janus achieves an effective trade-off between scalability and efficiency, making it practical for large-scale wireless network digital twin construction across different LLMs.
% we observe that Janus achieves acceptable wall-clock execution time of up to 6.72 s, 16.18 s, and 17.19 s at a network size of 200 for GPT, Claude, and DeepSeek, respectively, with corresponding token consumption of 4626.65, 4439.20, and 4012.88.
%These findings highlight the effectiveness and applicability of Janus for wireless network digital twin construction and its strong generalization across different LLMs.

$\bullet$ \textbf{Network orchestration}. 
We further evaluate Janus on two representative and more complex use cases, %We then consider two more complex use cases, 
as described in Sec. \ref{sec:sec:setup}: user association in a 50-node indoor hybrid RF/VLC network and routing in a OneWeb-constellation-based SAGIN. The performance and token consumption results for these two use cases are presented in Figs. \ref{fig:use_case_1} and \ref{fig:use_case_2}, respectively. Despite the significantly increased network complexity, Janus consistently achieves comparable performance to the Optimal benchmark. Specifically, the performance gap between Janus and Optimal is at most 5.42\% and 10.53\% for the user association and routing tasks, respectively. Meanwhile, the average token consumption remains modest, at only 2358 and 2649, respectively.
These results further demonstrate that Janus can effectively handle complex network orchestration tasks while maintaining both high performance and computational efficiency. Moreover, these results indicate the strong generalizability of Janus, enabling it to adapt to diverse network settings and heterogeneous wireless networking tasks using a unified framework.

\vspace{-3mm}
\section{Conclusion}
\vspace{-2mm}
\label{sec:conclusion}
\noindent In this paper, we presented \textbf{Janus}, an agentic LLM-based framework that simultaneously performs DT construction and network orchestration in dynamic wireless networks. Motivated by the limitations of LLM inference for graph-structured data and the strong dependence on prompt design, we introduced a \emph{one-agent, two-role} paradigm that integrates network modeling and decision-making within a closed-loop system. By designing the agentic NetTwin to integrate planner, modeler, and coder roles with a unified prompt module, Janus enables structured reasoning, iterative refinement, and task-aware interaction.
Extensive evaluations demonstrate that Janus achieves accurate digital twin construction while maintaining reasonable construction time. Moreover, it achieves near-optimal performance across complex network orchestration tasks with moderate computational overhead, highlighting its effectiveness and generalizability.
Janus represents a step toward fully autonomous, AI-native wireless network systems.

%\section{Acknowledgments}
\begin{spacing}{1}
\bibliographystyle{ieeetr}

\bibliography{ref}

\begin{thebibliography}{10}
\providecommand{\url}[1]{#1}
\csname url@samestyle\endcsname
\providecommand{\newblock}{\relax}
\providecommand{\bibinfo}[2]{#2}
\providecommand{\BIBentrySTDinterwordspacing}{\spaceskip=0pt\relax}
\providecommand{\BIBentryALTinterwordstretchfactor}{4}
\providecommand{\BIBentryALTinterwordspacing}{\spaceskip=\fontdimen2\font plus
\BIBentryALTinterwordstretchfactor\fontdimen3\font minus \fontdimen4\font\relax}
\providecommand{\BIBforeignlanguage}[2]{{%
\expandafter\ifx\csname l@#1\endcsname\relax
\typeout{** WARNING: IEEEtran.bst: No hyphenation pattern has been}%
\typeout{** loaded for the language `#1'. Using the pattern for}%
\typeout{** the default language instead.}%
\else
\language=\csname l@#1\endcsname
\fi
#2}}
\providecommand{\BIBdecl}{\relax}
\BIBdecl

\bibitem{10855897}
F.~Khoramnejad and E.~Hossain, ``{Generative AI for the Optimization of Next-Generation Wireless Networks: Basics, State-of-the-Art, and Open Challenges},'' \emph{IEEE Communications Surveys \& Tutorials}, vol.~27, no.~6, pp. 3483--3525, 2025.

\bibitem{xu2024generate}
Y.~Xu, S.~He, J.~Chen, Z.~Wang, Y.~Song, H.~Tong, G.~Liu, J.~Zhao, and K.~Liu, ``{Generate-on-Graph: Treat LLM as Both agent and KG for Incomplete Knowledge Graph Question Answering},'' in \emph{Proceedings of the 2024 Conference on Empirical Methods in Natural Language Processing}, 2024, pp. 18\,410--18\,430.

\bibitem{11477808}
B.~Wei, R.~Jiang, R.~Zhang, Y.~Liu, D.~Niyato, Y.~Sun, Y.~Lu, Y.~Li, S.~Mao, C.~Yuen, M.~D. Renzo, and M.~Peng, ``{Large Language Models for Optimization in Next-Generation Wireless Network Management: A Survey},'' \emph{IEEE Communications Surveys \& Tutorials}, vol.~28, pp. 5713--5746, 2026.

\bibitem{zhou2024large}
H.~Zhou, C.~Hu, D.~Yuan, Y.~Yuan, D.~Wu, X.~Liu, and C.~Zhang, ``{Large Language Model (LLM)-Enabled in-Context Learning for Wireless Network Optimization: A Case Study of Power Control},'' \emph{arXiv preprint arXiv:2408.00214}, 2024.

\bibitem{11449190}
T.~C. Dao, N.~C. Luong, X.~Li, T.~Huynh-The, D.~Niyato, and D.~I. Kim, ``{Autonomous Algorithm Synthesis via Large Language Models for Multi-Hop Routing in IoT Networks},'' \emph{IEEE Wireless Communications Letters}, pp. 1--1, 2026.

\bibitem{hong2024llm}
Y.~Hong, J.~Wu, and R.~Morello, ``{LLM-Twin: Mini-Giant Model-driven beyond 5G Digital Twin Networking Framework with Semantic Secure Communication and Computation},'' \emph{Scientific Reports}, vol.~14, no.~1, p. 19065, 2024.

\bibitem{10592370}
H.~Du, G.~Liu, Y.~Lin, D.~Niyato, J.~Kang, Z.~Xiong, and D.~I. Kim, ``{Mixture of Experts for Intelligent Networks: A Large Language Model-enabled Approach},'' in \emph{2024 International Wireless Communications and Mobile Computing (IWCMC)}, 2024, pp. 531--536.

\bibitem{10829752}
Y.~Song, Y.~Zhang, A.~Zhou, Y.~Shi, S.~Shen, X.~Tang, J.~Li, M.~Zhang, and D.~Wang, ``{Synergistic Interplay of Large Language Model and Digital Twin for Autonomous Optical Networks: Field Demonstrations},'' \emph{IEEE Communications Magazine}, vol.~63, no.~6, pp. 90--96, 2025.

\bibitem{10757470}
S.~Jiang, B.~Lin, Y.~Wu, and Y.~Gao, ``{LINKs: Large Language Model Integrated Management for 6G Empowered Digital Twin NetworKs},'' in \emph{2024 IEEE 100th Vehicular Technology Conference}, 2024, pp. 1--6.

\bibitem{duran2025gentwin}
K.~Duran, H.~Shin, T.~Q. Duong, and B.~Canberk, ``{GenTwin: Generative AI-Powered Digital Twinning for Adaptive Management in IoT Networks},'' \emph{IEEE Transactions on Cognitive Communications and Networking}, vol.~11, no.~2, pp. 1053--1063, 2025.

\bibitem{bao2025dynamic}
R.~Bao, N.~Xue, Y.~Sun, and Z.~Chen, ``{Dynamic Quality-Latency Aware Routing for LLM Inference in Wireless Edge-Device Networks},'' in \emph{2025 IEEE/CIC International Conference on Communications in China}, 2025, pp. 1--6.

\bibitem{11358889}
W.~Lee and J.~Park, ``{LLM-Empowered Resource Allocation in Wireless Communications Systems},'' \emph{IEEE Access}, vol.~14, pp. 15\,260--15\,272, 2026.

\bibitem{10901312}
H.~Zhang, A.~Bin~Sediq, A.~Afana, and M.~Erol-Kantarci, ``{Large Language Models in Wireless Application Design: In-Context Learning-enhanced Automatic Network Intrusion Detection},'' in \emph{2024 IEEE Global Communications Conference}, 2024, pp. 2479--2484.

\bibitem{11071329}
X.~Liu, S.~Gao, B.~Liu, X.~Cheng, and L.~Yang, ``Llm4wm: Adapting llm for wireless multi-tasking,'' \emph{IEEE Transactions on Machine Learning in Communications and Networking}, vol.~3, pp. 835--847, 2025.

\bibitem{10.1145/3718958.3750505}
C.~Wang, X.~Zhang, R.~Lu, X.~Lin, X.~Zeng, X.~Zhang, Z.~An, G.~Wu, J.~Gao, C.~Tian, G.~Chen, G.~Liu, Y.~Liao, T.~Lin, D.~Cai, and E.~Zhai, ``{Towards LLM-Based Failure Localization in Production-Scale Networks},'' in \emph{Proceedings of the ACM SIGCOMM 2025 Conference}, 2025, p. 496–511.

\bibitem{guo2023gpt4graph}
J.~Guo, L.~Du, H.~Liu, M.~Zhou, X.~He, and S.~Han, ``{Gpt4graph: Can Large Language Models Understand Graph Structured Data? An Empirical Evaluation and Benchmarking},'' \emph{arXiv preprint arXiv:2305.15066}, 2023.

\bibitem{sahoo2024systematic}
P.~Sahoo, A.~K. Singh, S.~Saha, V.~Jain, S.~Mondal, and A.~Chadha, ``{A Systematic Survey of Prompt Engineering in Large Language Models: Techniques and Applications},'' \emph{arXiv:2402.07927}, 2024.

\bibitem{jiang2025llmopt}
C.~Jiang, X.~Shu, H.~Qian, X.~Lu, J.~Zhou, A.~Zhou, and Y.~Yu, ``{LLMOPT: Learning to Define and Solve General Optimization Problems from Scratch},'' in \emph{Proceedings of the 13th International Conference on Learning Representations (ICLR)}, Singapore, Singapore, 2025.

\bibitem{eutelsat}
{Eutelsat OneWeb LEO Constellation }. \url{https://www.eutelsat.com/satellite-network/oneweb-leo-constellation}.

\end{thebibliography}
\end{spacing}
\end{document}